\documentclass[aip,jcp,reprint,longbibliography]{revtex4-2}
\usepackage{epsfig}
\usepackage{color}
\usepackage{amsmath}
\usepackage{amsfonts}
\usepackage{natbib}
\usepackage{notes2bib}

\newcommand{\degree}{^{\circ} }

\newcommand{\avg}[1]{\left<#1\right>}

\newcommand{\vect}[1]{\mathbf{#1}}

\DeclareMathOperator{\Tr}{Tr}

\newcommand{\ie}{\emph{i.e.}}

\newcommand{\etal}{\emph{et al.}}

\newcommand{\kT}{\ensuremath{k_{\rm B}T}}

\newcommand{\Hb}{\ensuremath{\mathcal{H}}}
\newcommand{\Zb}{\ensuremath{\mathcal{Z}}}

\newcommand{\chn}{{C-H$\cdots$N }} 
\newcommand{\nhpi}{{N-H$\cdots\pi$ }} 
\newcommand{\aacrystal}{{acetylene:ammonia (1:1)}}

\begin{document}


\title{Nuclear Quantum Effects in the Acetylene:Ammonia Plastic Co-crystal}


\author{Atul C. Thakur}
\author{Richard C. Remsing}
\email[]{rick.remsing@rutgers.edu}
\affiliation{Department of Chemistry and Chemical Biology, Rutgers University, Piscataway, NJ 08854}




\begin{abstract}
Organic molecular solids can exhibit rich phase diagrams. 
In addition to structurally unique phases, translational and rotational degrees of freedom can melt at different
state points, giving rise to partially disordered solid phases.
The structural and dynamic disorder in these materials can have a significant impact on the physical properties of the organic solid,
necessitating a thorough understanding of disorder at the atomic scale. 
When these disordered phases form at low temperatures, especially in crystals with light nuclei, 
the prediction of materials properties can be complicated by the importance of nuclear quantum effects. 
As an example, we investigate nuclear quantum effects on the structure and dynamics of 
the orientationally-disordered, translationally-ordered plastic phase of the \aacrystal~co-crystal that is expected to exist on the surface of Saturn's moon Titan. 
Titan's low surface temperature ($\sim 90$~K) suggests that the quantum mechanical behavior of nuclei may be important in this and other molecular solids in these environments.
By using neural network potentials combined with ring polymer molecular dynamics simulations,
we show that nuclear quantum effects increase orientational disorder and rotational dynamics within the \aacrystal~co-crystal by weakening hydrogen bonds. 
Our results suggest that nuclear quantum effects are important to accurately model molecular solids and their physical properties in low temperature environments.
\end{abstract}


\maketitle

\raggedbottom

\section{Introduction}

Molecular crystals play important roles in a wide variety of fields, including pharmaceuticals~\cite{beran2016modeling, kumar2018approaches}, semiconductor technologies~\cite{coropceanu2007charge, zhang2018light, bronstein2020role}, energy storage and transport~\cite{hirshberg2014calculations, zhang2020aromatic, liu2020novel}, pesticides~\cite{zhu2021imidacloprid}, and atmospheric and planetary sciences~\cite{thakur2023molecular, czaplinski2023experimental,cable2021titan,maynard2018prospects, maynard2016co}.   
Like the more familiar class of atomic crystals, molecular crystals are crystalline solids made of repeating molecular subunits. 
When these solids are made of two or more types of molecules, they are typically called molecular co-crystals~\cite{kirchner2010co, kumar2018approaches, cable2021titan}. 
Molecular co-crystals often exhibit thermodynamic and mechanical properties that are different from the corresponding single component solids, offering unique opportunities for tunable materials~\cite{sun2019cocrystal,karki2009improving,dai2018pharmaceutical}.  
Molecular crystals are usually held together by weak interactions like hydrogen bonding and dispersion interactions. 
As a result, molecular crystals often display rich phase behavior.  
In addition to the usual translationally and orientationally ordered crystalline solid phase, translationally ordered and orientationally disordered (or vice versa) phases can also appear. 
Solids that exhibit translational order but lack orientational order are known as plastic crystals~\cite{klein1990simulation, klein1985computer, bounds1980molecular, nose1983structural}.   
The plasticity, as suggested by the name, refers to the additional mechanical flexibility endowed by the presence of orientational disorder.    
Plastic phases can also exhibit different chemical and physical properties than the corresponding orientationally ordered phases.
For instance, some plastic phases exhibit different thermodynamic and electrical properties than the ordered phase~\cite{lynden1994translation}.
Importantly, these differences in physical properties can lead to significant macroscopic consequences.

Many molecular co-crystals are predicted to form on the cold ($\sim$90~K) surface of Saturn's moon Titan~\cite{cable2021titan, maynard2018prospects, maynard2016co, czaplinski2023experimental, cable2019co, cable2018acetylene, cable2020properties, vu2014formation, maynard2016co, vu2022Buta}. 
As a result, molecular crystals and co-crystals on Titan play an analogous role to minerals on Earth and have been termed cryominerals.
On Titan and similar planetary bodies, the changes in physical properties arising from plastic crystal phases could have implications for geochemical processes
and even surface geology~\cite{thakur2023molecular, cable2021titan,maynard2018prospects, maynard2016co}.
To begin to understand plastic phases of these cryominerals, we use molecular dynamics (MD) simulations to investigate the \aacrystal~plastic co-crystal~\cite{thakur2023molecular, cable2021titan, cable2018acetylene}.
In particular, the quantum mechanical nature of atomic nuclei is expected to be important at the cold temperatures on Titan's surface,
and we focus on quantifying the impact of nuclear quantum effects on the structure and dynamics of the \aacrystal~plastic co-crystal.
Nuclear quantum effects (NQEs) on structure and dynamics are significant for light nuclei like hydrogen even at room temperature (300~K)~\cite{markland2018nuclear, ceriotti2016nuclear, remsing2023modeling}.
As a result, the common approximation of using classical nuclei in molecular simulations may be inadequate for many molecular crystals.
Indeed, NQEs can play important roles in determining thermodynamic stability~\cite{rossi2016anharmonic, kapil2022complete}, thermal conductivity~\cite{rossi2016anharmonic}, expansivity, and mechanical properties~\cite{ko2018thermal}. 
At the cryogenic conditions on Titan, NQEs are anticipated to play an even greater role in determining the properties of molecular co-crystals.
Here, we quantify the role of NQEs in determining the structure and dynamics of the \aacrystal~co-crystal using neural network potential-based thermostatted ring polymer MD simulations. 
We first discuss simulation details and the theory of approximate real time dynamics within the ring polymer molecular dynamics framework. 
Then, we discuss the translational and orientational structure of the \aacrystal~co-crystal as obtained from the classical and quantum simulations, 
where classical and quantum refer to the treatment of nuclear degrees of freedom, and show that NQEs increase rotational disorder within the co-crystal.
We then quantify NQEs on dynamics within the co-crystal through examination of vibrational densities of states and rotational time correlation functions,
and we connect the observed increase in rotational diffusion by NQEs to a similar increase in hydrogen bond dynamics and a lowering of rotational free energy barriers. 
Our results suggest that NQEs can be significant in plastic crystals, especially at low temperatures, and should be taken into account when modeling these disordered solids. 
%

\section{Simulation Details}

We performed neural network (NN)-based MD simulations of the \aacrystal~co-crystal using the LAMMPS software package~\cite{thompson2022lammps}. 
We trained a neural equivariant interatomic potential (NequIP)~\cite{batzner20223} on ab initio density functional theory (DFT) data generated using CP2K~\cite{kuhne2020cp2k, hutter2014cp2k, vandevondele2003efficient}. 
We also trained a deep potential (DeePMD) model~\cite{zhang2018end,lu202186,zeng2023deepmd,wang2018deepmd} on a larger data set, but this potential resulted in inaccuracies in dynamical properties, as discussed further in the Appendix.
The training data was obtained from the DFT-based ab initio molecular dynamics (AIMD) simulations~\cite{vandevondele2005quickstep} at 90~K and 600~K.  
The ab initio energy and forces were computed using the Perdew-Burke-Ernzerhof (PBE) treatment of exchange-correlation (XC) interactions~\cite{perdew1996generalized} along with Grimme's D3 correction for long range dispersion interactions~\cite{grimme2010consistent,grimme2011effect}.
We used molecularly optimized (MOLOPT) G\"{o}decker-Teter-Hutter (GTH) triple-$\zeta$ single polarization (TZVP-MOLOPT-GTH) basis sets along with G\"{o}decker-Teter-Hutter (GTH) relativistic, norm-conserving, separable, dual-space pseudopotentials~\cite{krack2005pseudopotentials,hartwigsen1998relativistic,goedecker1996separable}. 
More details on the DFT simulations can be found in our earlier work~\cite{thakur2023molecular}. 
The NN potential was constructed using the NequIP code to optimize the force and energy loss terms according to~\cite{batzner20223}
\begin{align}
\mathcal{L} = \lambda_E \left| \left| \hat{E} - E \right| \right|^2 +  \frac{\lambda_F}{3N} \sum_{i=1}^{N} \sum_{\alpha=1}^{3}  \left| \left|  -\frac{\partial \hat{E}}{\partial r_{i, \alpha}} - F_{i, \alpha} \right| \right|,
\end{align}
where $N$ refers to the total number of atoms in the system, $\alpha$ indicates each independent Cartesian direction, $\hat{E}$ is the predicted potential energy and $\lambda$ refers to the weights for the energies ($\lambda_E$) and forces ($ \lambda_F$).  
We fixed the maximum rotation order for the network's features at $l = 1$ and allowed the features with odd parity to be included with the network parity turned on.
The locality for the features was imposed with a radial cutoff of 6~\AA.
We uniformly sampled the DFT dataset to select 2000 configurations, half from simulations at 90~K and half from simulations at 600~K.   
From the total data set, a validation set of 400 snapshots was picked randomly. 
The energy data was internally shifted for each particle with per atom total energy mean while forces were rescaled with force root mean square deviation. 
Within the network, we used four interaction blocks with an initial learning rate of 0.005 and a batch size of 5.
The learning rate gets rescaled by a factor of 0.5 if the validation loss of atomic forces has not improved over 100 consecutive epochs.  
We terminated the training process after around 4000 epochs because the errors in the forces and energies were sufficiently small to give good agreement with the AIMD simulations. 
The training process continuously updates the model with the best force validation loss and we froze the model with the overall best validation loss to be used for MD simulations.  
We performed classical MD simulations of the \aacrystal~co-crystal using the above-described NequIP NN potential.
All simulations were performed using LAMMPS with the nequip pair style provided by the pair-nequip library~\cite{batzner20223}.  
The crystal structure for the \aacrystal~co-crystal was taken from the experimental crystal structure reported by Boese~\etal \ (Cambridge Structural Database ID: FOZHOS)~\cite{boese2009synthesis}. 
We used a $4 \times 4 \times 4$ supercell consisting of 1024 atoms; 128 acetylene and 128 ammonia molecules.  
The system was equilibrated at 90~K for at least 500~ps in the canonical (NVT) ensemble using a Nos\'{e}-Hoover thermostat with three thermostat chains~\cite{nose1984unified, nose1984molecular,hoover1985canonical,martyna1992nose}.
We also performed simulations with classical nuclei at 30~K, where the crystal is not plastic and is orientationally ordered, to contrast with the plastic phase and highlight the effects of orientational disorder.
For simulations at 30~K, the temperature was scaled down gradually starting from an equilibrated configuration at 90~K using Nos\'{e}-Hoover chains.
To gather accurate dynamics, we sampled the system in the microcanoncal (NVE) ensemble for at least 1.5~ns at 90~K and for 500~ps at 30~K. 
The equations of motion were integrated using the velocity Verlet algorithm as implemented in LAMMPS with a timestep of 1~fs.  
To quantify NQEs, we performed NN potential-based thermostatted ring polymer molecular dynamics (TRPMD) simulations using LAMMPS interfaced with i-PI~\cite{kapil2019pi}.
The i-PI package exploits a client-server interface in which the integration of the equations of motion for the ring polymer is handled by i-PI while the computationally intensive energy and force calculations are offloaded to client codes. 
We used the LAMMPS client interface along with the NequIP neural network potential for evaluation of energies and forces.
Convergence of the ring polymer discretization with respect to Trotter bead number was evaluated using NN-based simulations of the \aacrystal~co-crystal and achieved at 128 beads at 90~K, in agreement with previous simulations of hydrocarbons at similar temperatures~\cite{uhl2016accelerated}. 
To avoid numerical instabilities and enable the use of a larger timestep of 0.5~fs in the TRPMD simulations, we used the Cayley integrator with the BCOCB factorization scheme~\cite{korol2019cayley, rosa2021generalized, korol2020dimension}.  
Our TRPMD~\cite{rossi2014remove,craig2004quantum} simulations employed the same $4\times4\times4$ simulation supercell as the classical simulations.
The starting configuration for the TRPMD simulations was chosen randomly from the classical simulation and was further equilibrated at 90~K in the canonical (NVT) ensemble with a global stochastic velocity rescaling (SVR) thermostat~\cite{bussi2009isothermal} for over 2~ps. 
To ensure ergodicity in the TRPMD simulations, we shoot two trajectories, each starting from a randomly selected state point along the NVT RPMD trajectory.  
Both TRPMD trajectories were sampled for 25~ps. 
The internal ring polymer modes were thermostatted during sampling using the white noise PILE-G thermostat, while all of the physical modes of the system remained unthermostatted.
This procedure is expected to be accurate for extracting approximate quantum dynamics while still mitigating ergodicity issues arising in completely unthermostatted RPMD simulations~\cite{rossi2014remove,craig2004quantum}.  
The results presented here reflect the average over both TRPMD simulations.
%

\section{Real time Dynamics with TRPMD}

We begin with a brief review of the definitions of various quantum time correlations functions and their corresponding RPMD approximations~\cite{craig2004quantum}. 
We use the Kubo-transformed definition of time correlation functions as our exact quantum correlations which we approximate by time correlations from RPMD simulations.
However, we note that Kubo transforms are not the only definition of exact quantum time correlations and symmetrized quantum correlations have also been proposed that are related to the Kubo-transformed definition~\cite{habershon2013ring, craig2004quantum}.

Translational dynamics can be probed using the classical velocity autocorrelation function 
\begin{equation}
C_v(t) = \frac{ \avg{ \vect{v}(0) \cdot \vect{v}(t)}} { \avg{ \vect{v}(0) \cdot \vect{v}(0)}},
\label{eq:cvt}
\end{equation}
where $\vect{v}(t)$ indicates the instantaneous velocity of an atom at time $t$ and $\avg{\cdots}$ indicates an ensemble average. 
The vibrational power spectrum, $C_v(\omega)$, is the Fourier transform of $C_v(t)$. 
The quantum mechanical Kubo-transformed velocity autocorrelation function is~\cite{miller2005quantum}
\begin{equation}
{C}_v(t) = \frac{1}{\beta}\int_{0}^{\beta} d\lambda \avg{\vect{v}(-i\lambda\hbar)\cdot\vect{v}(t)},
\end{equation}
where $\beta=1/\kT$ is the inverse Boltzmann temperature and $\vect{v}(t)$ refers to the Heisenberg evolved velocity operator of an atom at time $t$.
The Kubo-transformed correlation function is the most classical version of an exact quantum mechanical correlation function that one can construct, and it obeys the same symmetries as the classical time correlation functions~\cite{habershon2013ring, craig2004quantum}.   
To compute quantum time correlation functions,
we use the RPMD approximation to the Kubo-transformed time correlation function.  
The RPMD approximation to the Kubo-transformed velocity autocorrelation function is given by~\cite{miller2005quantum}
\begin{equation}
\begin{aligned}
{C}_v(t) &\simeq \frac{1}{(2\pi\hbar)^{9NP}\Zb_P} \int\int \prod_{k=1}^{3N} \prod_{\alpha=1}^{P} d\vect{p}_k^{(\alpha)} d\vect{q}_k^{(\alpha)} \\
&\times e^{-\beta_P \Hb_P(\{\vect{p}_k^{(\alpha)}\},\{\vect{q}_k^{(\alpha)}\})}  \vect{ \overline{v} }(0) \cdot \vect{ \overline{v} }(t),
\end{aligned}
\end{equation}
where $\beta_P = \beta/P$, $\Hb_P$ is the ring polymer Hamiltonian, $ \vect{q}^{(\alpha)}$ and $ \vect{p}^{(\alpha)}$ refer to the position and momentum coordinates of bead index $\alpha$, and $\Zb_P$ is the canonical partition function of the path integral ring polymer system with $P$ beads~\cite{miller2005quantum}.
The operator $\vect{ \overline{v} }(t)$ is the instantaneous bead-averaged velocity operator of an atom at time $t$ given by 
\begin{equation}
\vect{ \overline{v} }(t) = \frac{1}{P} \sum\limits_{\alpha=1}^{P}  \vect{v}^{(\alpha)}(t).
\end{equation} 
Note that the RPMD approximation to the Kubo-transformed quantum correlation function becomes exact in the short-time limit ($t \rightarrow 0$) and coincides with the classical correlation function in the limit that the ring-polymer collapses onto a single bead ($P \rightarrow 1$). 
The rotational dynamics of molecules can be quantified using rotational (or orientational) time correlation functions~\cite{TheorySimpLiqs}. 
For classical nuclei, the rotational time correlation functions we examine are of the form
\begin{equation}
C_l(t) =  \avg{P_l(\vect{\hat{n}}(0) \cdot \vect{\hat{n}}(t))},
\end{equation}
where $\vect{\hat{n}}(t)$ is the instantaneous unit vector pointing along the N-H bond for ammonia molecules
or the C-C bond for acetylene molecules, and $P_l(x)$ is the $l$th Legendre polynomial. 
We quantify rotational motion with $C_1(t)$ and $C_2(t)$, which correspond to the first and second order
Legendre polynomials.
These correlation functions can be experimentally measured through IR and Raman spectroscopies, respectively~\cite{thakur2021distributed, gordon1965molecular,gordon1965relations}.
The classical expression for $C_1(t)$ is
\begin{equation}
C_1(t) =  \avg{\vect{\hat{n}}(0) \cdot \vect{\hat{n}}(t)},
\end{equation}
and $C_2(t)$ is given by
\begin{equation}
C_2(t) = \avg{\frac{3}{2}(\vect{\hat{n}}(0) \cdot \vect{\hat{n}}(t))^2 - \frac{1}{2}}.
\end{equation}
The corresponding generalized Kubo-transformed expression for the $l$th order quantum rotational time correlation function is~\cite{miller2005quantum}
\begin{equation}
{C}_l(t) = \int_{0}^{\beta} d\lambda \avg{P_l(\vect{\hat{n}}(-i\lambda\hbar) \cdot \vect{\hat{n}}(t))}.
\end{equation}
Therefore, the $l=1$ Kubo-transformed rotational time correlation function is
\begin{equation}
{C}_1(t) = \int_{0}^{\beta} d\lambda  \avg{\vect{\hat{n}}(-i\lambda\hbar) \cdot \vect{\hat{n}}(t)},
\end{equation}
and the corresponding $l=2$ expression is
\begin{equation}
{C}_2(t) = \int_{0}^{\beta} d\lambda \avg{\frac{3}{2}(\vect{\hat{n}}(-i\lambda\hbar) \cdot \vect{\hat{n}}(t))^2 - \frac{1}{2}}.
\end{equation}
The RPMD approximation to the Kubo-transformed $l=1$ rotational correlation can then be written as~\cite{miller2005quantum, wilkins2017nuclear}
\begin{equation}
\begin{aligned}
{C}_1(t) &\simeq \frac{1}{(2\pi\hbar)^{9NP}\Zb_P} \int\int \prod_{k=1}^{3N} \prod_{\alpha=1}^{P} d\vect{p}_k^{(\alpha)} d\vect{q}_k^{(\alpha)} \\
&\times e^{-\beta_P \Hb_P(\{\vect{p}_k^{(\alpha)}\},\{\vect{q}_k^{(\alpha)}\})}   \vect{\overline{n}}(0) \cdot \vect{\overline{n}}(t),
\end{aligned}
\end{equation}
where $\vect{\overline{n}}$ is a bead-averaged operator,
\begin{equation}
\vect{\overline{n}}(t) = \frac{1}{P} \sum\limits_{\alpha=1}^{P}  \vect{\hat{n}}^{(\alpha)}(t).
\end{equation} 
The RPMD approximation to the $l=2$ rotational time correlation function is given by~\cite{miller2005quantum, wilkins2017nuclear}
\begin{equation}
\begin{aligned}
{C}_2(t) &\simeq \frac{3}{2(2\pi\hbar)^{9NP}\Zb_P} \int\int \prod_{k=1}^{3N} \prod_{\alpha=1}^{P} d\vect{p}_k^{(\alpha)} d\vect{q}_k^{(\alpha)} \\
&\times e^{-\beta_P \Hb_P(\{\vect{p}_k^{(\alpha)}\},\{\vect{q}_k^{(\alpha)}\})}  \Tr[ \vect{ \overline{M} }(0)\vect{ \overline{M} }(t)] - \frac{1}{2},
\end{aligned}
\end{equation}
where  $\Tr$ indicates the trace of a matrix and $\vect{ \overline{M} }(t)$ is a bead-averaged third rank tensor,
\begin{equation}
\vect{ \overline{M} }(t) = \frac{1}{P} \sum\limits_{\alpha=1}^{P}  \vect{\hat{n}}^{(\alpha)}(t)\vect{\hat{n}}^{(\alpha)}(t)^T.
\end{equation} 
The operator $\vect{\hat{n}}^{(\alpha)}(t)$ is a $ 3 \times 1$ dimensional vector at real time $t$ and at imaginary time (or bead index) $\alpha$. 

\begin{figure}[tb]
\begin{center}
\includegraphics[width=0.4\textwidth]{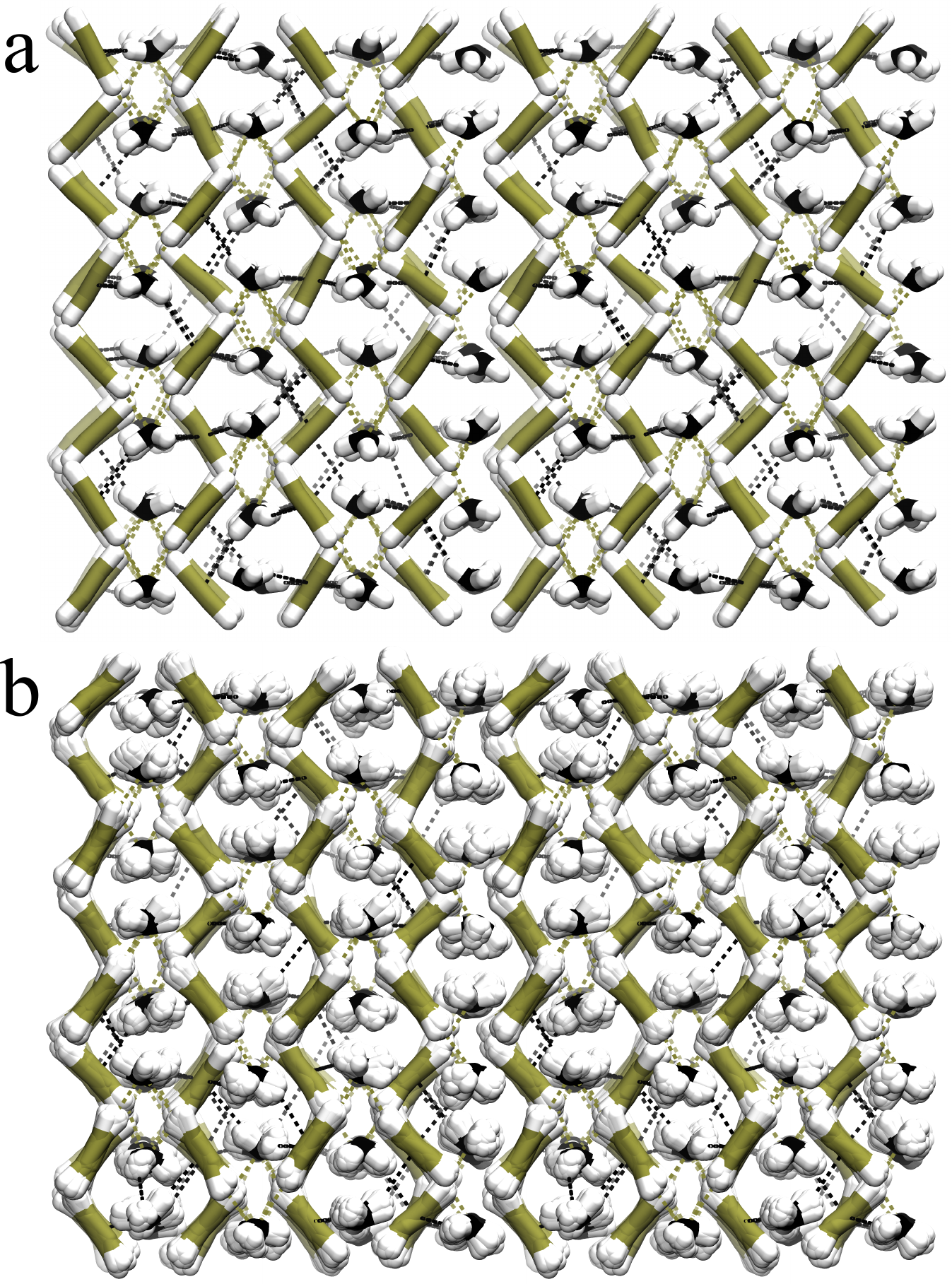}
\end{center}
\caption
{
Snapshots showing the zig-zag structure of the \aacrystal~co-crystal from (a) classical NN based simulations and (b) quantum TRPMD simulations, where every fifth bead in the ring polymer is shown. Carbon atoms are colored gold, nitrogen atoms are colored black, and hydrogen atoms are colored white. Both \chn and \nhpi hydrogen bonds are shown and colored according to the respective donor atoms. Both panels depict views from the crystal $z$-axis. 
}
\label{fig:snapshots}
\end{figure}

%
We also quantify hydrogen bond dynamics with the time correlation function~\cite{luzar1996effect, luzar1996hydrogen, luzar2000resolving}
\begin{equation}
\tilde{C}_{\rm{HB}}(t) = \avg{h(0)h(t)},
\end{equation}
where the hydrogen bond indicator function is given by $h(t) = \Theta[R_c - R(t)]\Theta[\phi_c - \phi(t)]$, $\Theta(x)$ is the Heaviside step function,
$R(t)$ is the distance between the two heavy atoms involved in the H-bond at time $t$,
and $\phi(t)$ is the donor-acceptor-H angle at time $t$.
The indicator function, $h(t)$, equals 1 if the H-bond distance and angle are less than the respective cutoffs
$R_c$ and $\phi_c$. 
We use $R_c=4$~\AA \ and $\phi_c=30\degree$ for both types of H-bonds.
The Kubo-transformed hydrogen bond time correlation function is 
\begin{equation}
\tilde{C}_{\rm HB}(t) = \frac{1}{\beta}\int_{0}^{\beta} d\lambda \avg{ h(-i\lambda \hbar) {h}(t) }.
\end{equation}
Then, the RPMD approximation to the quantum hydrogen bond correlation function is given by
\begin{equation}
\begin{aligned}
\tilde{C}_{\text{HB}}(t) &\simeq \frac{1}{(2\pi\hbar)^{9NP}\Zb_P} \int\int \prod_{k=1}^{3N} \prod_{\alpha=1}^{P} d\vect{p}_k^{(\alpha)} d\vect{q}_k^{(\alpha)} \\
&\times e^{-\beta_P\Hb_P(\{\vect{p}_k^{(\alpha)}\},\{\vect{q}_k^{(\alpha)}\})}  \overline{h}(0) \overline{h}(t), 
\end{aligned}
\end{equation}
where $\overline{h}(t)$ is a bead-averaged indicator function,
\begin{equation}
\overline{h}(t) = \frac{1}{P} \sum\limits_{\alpha=1}^{P}  \overline{h}^{(\alpha)}(t).
\end{equation} 
We normalize the hydrogen bond time correlation functions according to
\begin{equation}
{C}_{\text{HB}}(t) = \frac{\tilde{C}_{\text{HB}}(t)}{\tilde{C}_{\text{HB}}(0)}.
\end{equation} 
In general, $\tilde{C}_{\rm HB}(0)$ is smaller in a quantum system than a classical one. 
The reduction of $\tilde{C}_{\rm HB}(t)$ at $t=0$ results from the diffuse nature of the quantum particle,
such that some beads are H-bonded while others are not. 
Normalization of the H-bonding correlation function allows us to better
compare H-bond dynamics in the quantum and classical systems. 
%

\section{Results and Discussion}

\subsection{Nuclear Quantum Effects on Translational and Orientational Structure}

The \aacrystal~co-crystal consists of antiparallel planes of acetylene and ammonia molecules arranged in a zig-zag fashion~\cite{thakur2023molecular, cable2018acetylene,preston2012formation,boese2009synthesis, cable2021titan}, as shown in Fig.~\ref{fig:snapshots}.
The zig-zag arrangement of the molecules within the co-crystal is primarily held together by two types of interactions:
non-directional van der Waals interactions and directional hydrogen bonding interactions.  
The \aacrystal~co-crystal exhibits \chn and \nhpi hydrogen bonds~\cite{thakur2023molecular, cable2018acetylene,preston2012formation,boese2009synthesis}.
\chn hydrogen bonds result from electrostatic attractions between acetylene hydrogens and lone pairs of nitrogen atoms, 
and \nhpi hydrogen bonds result from electrostatic attractions between ammonia hydrogens and $\pi$ systems of acetylene molecules.
\nhpi hydrogen bonds are weaker than \chn hydrogen bonds. 
Thermal fluctuations at 90 K can disrupt the \nhpi hydrogen bonds, causing the ammonia molecules to be transiently free~\cite{thakur2023molecular}. 
Additionally, the ammonia molecules within the co-crystal are situated at positions of higher symmetry than the $C_3$ symmetry inherent to the ammonia molecules. 
An ammonia molecule cannot hydrogen bond to all six neighbors at once, and the presence of symmetry equivalent orientations can lead to orientational disorder.
%

\begin{figure}[tb]
\begin{center}
\includegraphics[width=0.30\textwidth]{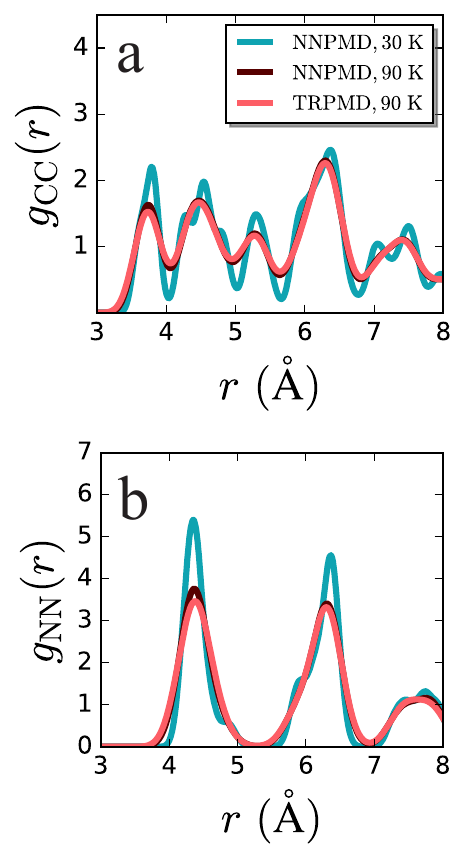}
\end{center}
\caption
{
Site-site radial distribution functions (RDFs), $g(r)$, computed for (a) acetylene carbons and (b) ammonia nitrogens within the classical (NNPMD) acetylene:ammonia (1:1) co-crystal at $T = 30$ and $90$~K. The results with the quantum nuclei at $T = 90$~K are also shown (TRPMD).
}
\label{fig:rdf1}
\end{figure}

%
We quantify the translational structure of the crystal by computing site-site radial distribution functions, $g_{\rm XY}(r)$, where ${\rm X}$ and ${\rm Y}$ indicate atomic or electronic sites. 
The classical C-C and N-N RDFs evaluated at 30~K are much sharper than the corresponding classical structures at 90~K, Fig.~\ref{fig:rdf1}.
This is expected due to thermal broadening at higher temperatures. 
At 90~K, NQEs have little impact on the C-C and N-N RDFs aside from a slight broadening of the peaks, Fig.~\ref{fig:rdf1}.
This suggests that NQEs do not significantly impact the structure of heavy atoms in the \aacrystal~co-crystal at 90~K.
%

\begin{figure}[tb]
\begin{center}
\includegraphics[width=0.48\textwidth]{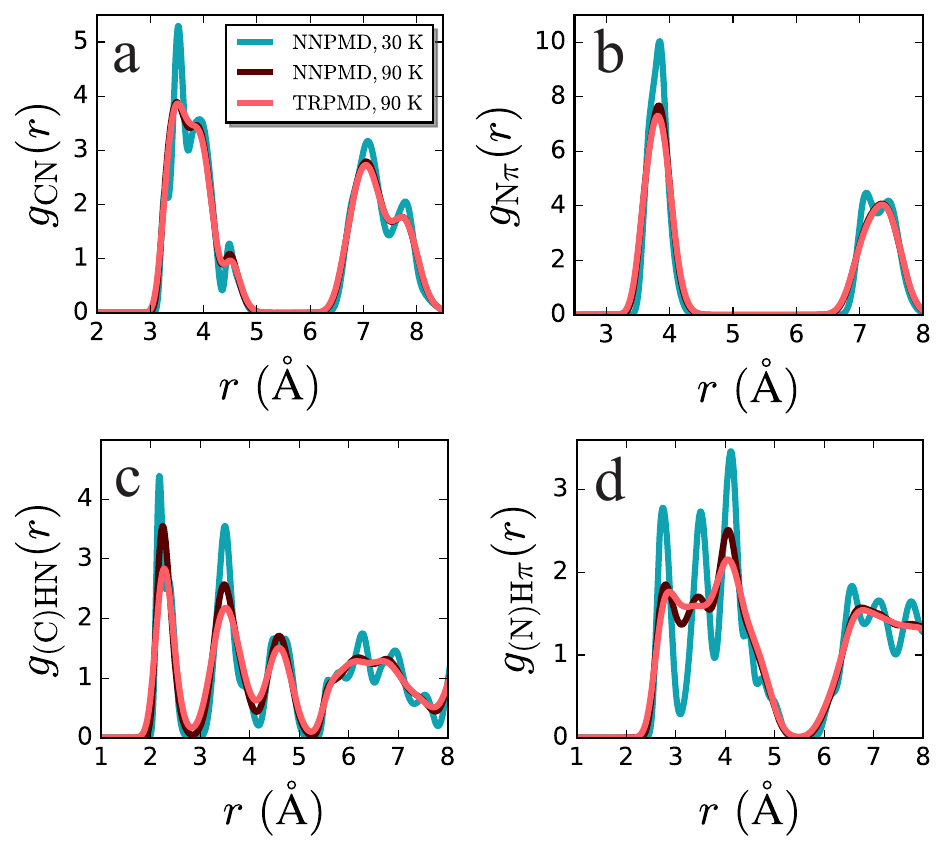}
\end{center}
\caption
{
Site-site radial distribution functions (RDFs), $g(r)$, relevant for sites forming (a, c) \chn~hydrogen bonds and (b, d) \nhpi~hydrogen bonds within the classical (NNPMD) acetylene:ammonia (1:1) co-crystal at $T = 30$ and $90$~K. The RDFs computed with quantum nuclei at $90$~K at are also shown (TRPMD).  
}
\label{fig:rdf2}
\end{figure}

%
We probe the hydrogen bonding structure within the co-crystal by computing the RDFs for the sites involved in hydrogen bonding interactions, Fig.~\ref{fig:rdf2}. 
For \chn hydrogen bonds, we compute the C-N and (C)H-N RDFs, where the notation (C)H-N indicates that the RDFs were computed between hydrogens of the acetylene molecules and the nitrogen atoms on the ammonia. 
Analogously, for \nhpi hydrogen bonds, we compute the N-$\pi$ and (N)H-$\pi$ structures, where the notation (N)H indicates the hydrogen atoms of the ammonia.
We define the $\pi$ site as the center of the C-C bond~\cite{thakur2023molecular}. 
The 30~K RDFs are sharper than those at 90~K, suggesting more structured hydrogen bonds within the crystal at low temperatures. 
Upon increasing the temperature from 30 to 90~K, 
the first three peaks in the (N)H-$\pi$ RDF broaden, hinting at the plastic crystal phase transition of the co-crystal. 
The C-N structure within the co-crystal is minimally perturbed with the inclusion of NQEs.  
However, the (C)H-N RDF slightly broadens upon inclusion of NQEs.
Similarly, the N-$\pi$ RDF remains relatively unaffected by NQEs, while the (N)H-$\pi$ RDF is slightly broadened. 
The peak broadening upon inclusion of NQEs suggests that the quantum hydrogen bond structure is more disordered than the corresponding classical network. 
This effect is primarily due to the hydrogen nuclei, which experience the largest NQEs due to their light mass and large thermal radius.
%

\begin{figure*}[tb]
\begin{center}
\includegraphics[width=0.98\textwidth]{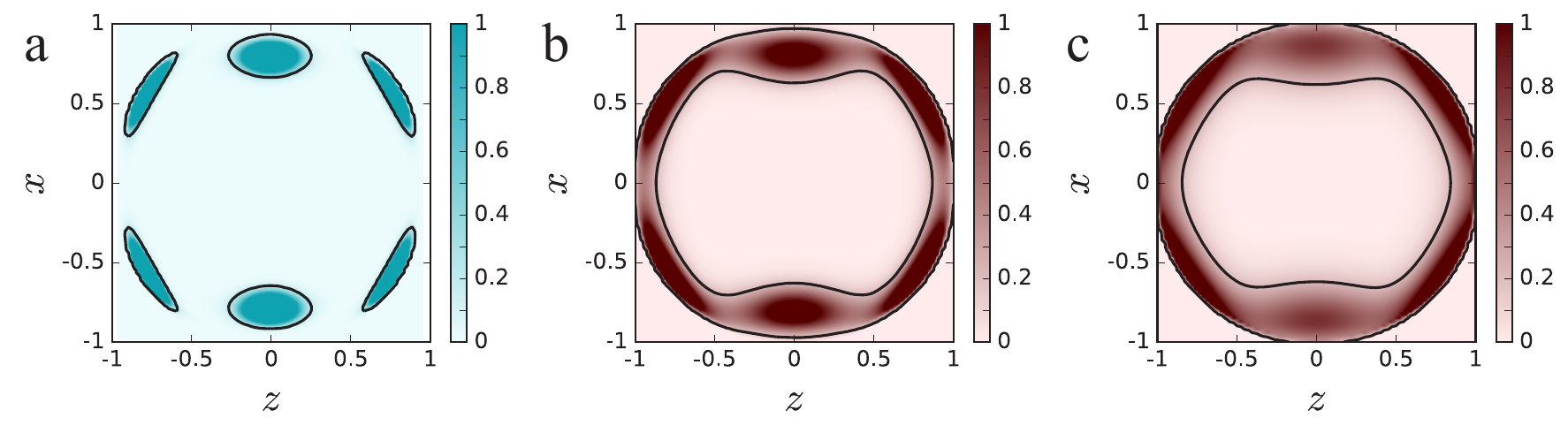}
\end{center}
\caption
{
Probability distributions of the ammonia H atom positions projected onto the crystal $xz$ plane
at (a) $T=30$~K and (b,c) $T=90$~K. 
Results for both (b) classical and (c) quantum nuclei are shown at 90~K. 
The contour line is drawn at a probability of 0.2. 
Peak heights are larger than one but the color scales are zoomed in to highlight the connectivity of the distributions at 90~K. 
}
\label{fig:os}
\end{figure*}

%
RDFs provide little information on the orientational order of the molecules within the co-crystal, although some information on the orientational order can be gleaned from the (N)H-$\pi$ correlations. 
We can better understand the orientational order of the molecules by projecting their positions on the crystallographic $xz$ plane, Fig.~\ref{fig:os}. 
The orientational distributions at 30~K show the presence of discrete disconnected regions of finite probability, indicating orientational order. 
In contrast, the continuous orientational probability distribution of the classical system at 90~K shows that ammonia molecules can adopt all angular positions in the $xz$ plane,
albeit some orientations are occupied more than others.
These preferred regions correspond to hydrogen bonded configurations with six distinct regions that arise from two different orientations of ammonia that hydrogen bond to neighboring acetylenes; these two orientations are discussed in more detail below. 
When NQEs are included, the probability distribution further broadens along the $xz$ plane and the ammonia molecules preferential orientation in specific directions is decreased.
The broadening of the probability distribution suggests that rotational disorder of ammonia molecules is even more pronounced in the quantum system. 
The continuous nature of the orientational distributions at 90~K suggests that ammonia molecules dynamically transition between different hydrogen bonded configurations
and that NQEs increase these transitions.
We quantify the classical and quantum dynamics associated with these transitions in the next section. 
%

\subsection{Nuclear Quantum Effects on Translational and Rotational Dynamics}

\begin{figure}[tb]
\begin{center}
\includegraphics[width=0.30\textwidth]{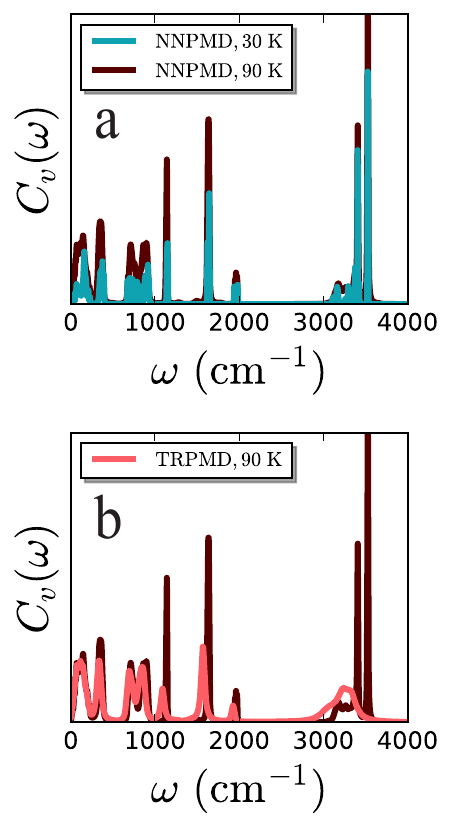}
\end{center}
\caption
{
Total all-atom vibrational power spectrum evaluated for acetylene:ammonia (1:1) co-crystal at (a) $T = 30$~K and (b) $T = 90$~K with (TRPMD) and without (NNPMD) nuclear quantum effects. 
}
\label{fig:pwrspec}
\end{figure}

%
To further understand the increased rotational disorder from NQEs, we investigate the dynamics within the crystal. 
We begin with a discussion of NQEs on vibrations within the co-crystal as quantified
by vibrational power spectra, Fig.~\ref{fig:pwrspec}.
The vibrational frequencies predicted are in good agreement with experimental results, as we showed previously~\cite{thakur2023molecular}. 
When heating the system from 30 to 90~K, the peaks slightly broaden and increase in intensity, reflecting increased fluctuations. 
The inclusion of NQEs significantly impacts the vibrational densities of states, especially the high frequency modes. 
The highest frequency peaks, resulting from C-H and N-H stretching modes, are broadened into a single wide peak.
Broadening of high frequency vibrational bands can be attributed to anharmonicity of the intermolecular vibrations, which becomes more important when zero point contributions are included~\cite{habershon2009competing}.   
In general, all vibrational peaks are red shifted and broadened by NQEs. 
The red shift with respect to the classical spectrum brings the vibrational frequencies in better agreement with experimental results~\cite{thakur2023molecular, cable2018acetylene, preston2012formation}. 
Red shifting of vibrational bands by NQEs also occurs in liquid water~\cite{paesani2006accurate} and has been rationalized as softening of the potential due to contributions from zero point fluctuations~\cite{paesani2007quantum}.

%

\begin{figure}[tb]
\begin{center}
\includegraphics[width=0.48\textwidth]{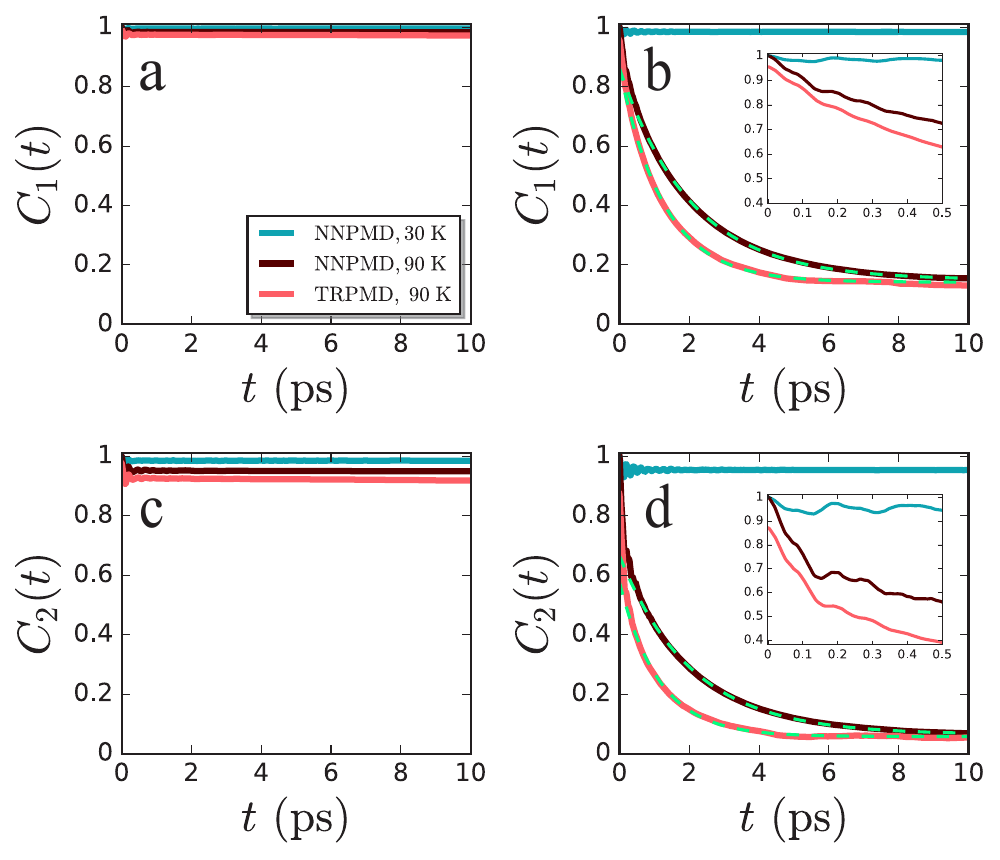}
\end{center}
\caption
{
Rotational time correlation functions, (a, b) $C_1(t)$ and (c, d) $C_2(t)$, evaluated with (TRPMD) and without (NNPMD) NQEs for (a, c) acetylene molecules and for (b, d) ammonia molecules within the \aacrystal~co-crystal. Results at both $T =  30$ and 90~K are shown. The dotted lines show the exponential fits to the correlation functions at long times. The insets in panels (b, d) shows the short time decay of ammonia rotational time correlation functions.  
}
\label{fig:rotcorrs}
\end{figure}

%
We now turn to investigating orientational dynamics within the \aacrystal~co-crystal.
For the acetylene molecules within the co-crystal, we compute $C_1(t)$ and $C_2(t)$ orientational time correlation functions for the C-C bond vector, Fig.~\ref{fig:rotcorrs}a,c.
For the ammonia molecules, the N-H bond vectors are used to define both orientational time correlation functions, Fig.~\ref{fig:rotcorrs}b,d.
For the acetylene molecules, both correlation functions show no appreciable long time decay.
In the short time limit, $t < 500$~fs, the classical correlations show a slight decay due to librational motion.
This minor decay is more apparent in the quantum correlation functions due to increased librational motion from zero-point fluctuations~\cite{wilkins2017nuclear}.
Finally, in the long time limit, both quantum and classical correlation functions saturate to a constant value.   
This overall lack of decay indicates that acetylene molecules do not undergo rotational diffusion and remain both translationally and orientationally ordered.

We apply a similar analysis to understand the orientational dynamics of ammonia molecules within the \aacrystal~co-crystal. 
Unlike the acetylene molecules, the $C_1(t)$ and $C_2(t)$ orientational correlations for the ammonia molecules exhibit an exponential decay at long times, indicating rotational diffusion.
Following our previous work~\cite{thakur2023molecular}, we estimate the correlation times $\tau_1$ and $\tau_2$ from $C_1(t)$ and $C_2(t)$ by numerically integrating the time correlation function
from 0 to 500~fs and integrating an exponential fit to the correlation function for longer times.
Inclusion of NQEs increases the rate of decay of these correlations and shortens the correlation times;
$\tau_1^{\rm c} = 1.5$~ps while $\tau_1^{\rm q} = 0.79$~ps and $\tau_2^{\rm c}=1.3$~ps while $\tau_2^{\rm q}=0.52$~ps, where c and q indicate classical and quantum correlation times, respectively.
At short times ($t < 500$~fs), $C_1(t)$ and $C_2(t)$ both exhibit rapid decay followed immediately by the onset of a characteristic glitch in the correlation functions (inset Fig.~\ref{fig:rotcorrs}b,d).
A similar glitch is also observed in the quantum time correlation functions and is suggestive of a librational rebound in cage-like environments~\cite{miller2005quantum}.  
In the diffusive regime ($t > 500$~fs), both classical and quantum correlations decay exponentially as the molecules eventually break through the cage overcoming the free energy barrier associated with rotations.
Overall, both classical and quantum correlation functions suggest that the ammonia molecules within the crystal exhibit a plastic phase, \ie~they remain translationally ordered but are orientationally disordered.  
Comparing the classical and quantum correlations for ammonia molecules at all times indicates that quantum correlations decay faster than the classical correlations.  
The rapid decay of quantum correlations is suggestive of increased orientational disorder of ammonia molecules within RPMD simulations and is a manifestation of NQEs.  
Another interesting NQE is the deviation of the value of both $C_1(t)$ and $C_2(t)$ at $t=0$ from unity. 
Such deviations from unity arise from quantum dispersions in the orientational coordinates~\cite{miller2005quantum}.  
We note that the RPMD approximation coincides with the exact Kubo-transformed quantum correlation function in the limit $t \rightarrow 0$, implying that quantum dispersions within the orientational coordinates is a purely quantum mechanical effect and not a consequence of the RPMD approximation.    
%

\subsection{Nuclear Quantum Effects on Hydrogen Bond Dynamics}

\begin{figure}[tb]
\begin{center}
\includegraphics[width=0.30\textwidth]{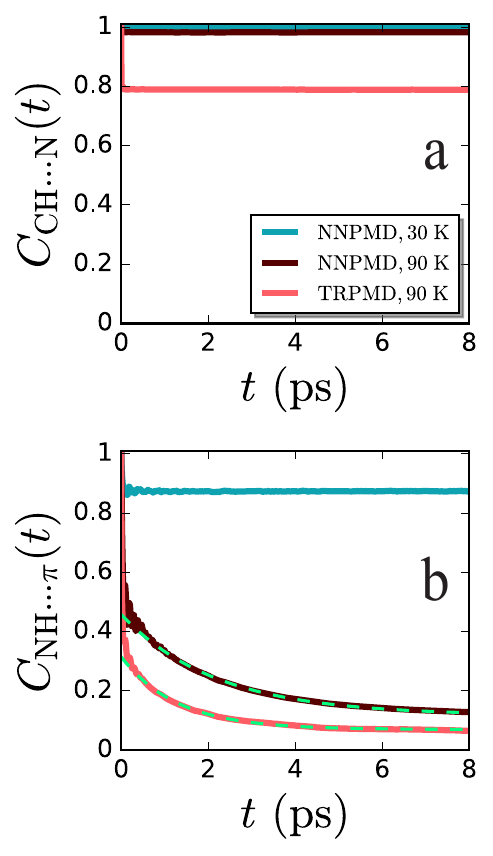}
\end{center}
\caption
{
Hydrogen bonding time correlation functions evaluated at $T = 30$ and 90~K for (a) \chn~and (b) \nhpi hydrogen bonds within the classical \aacrystal~co-crystal (NNPMD). The hydrogen bond dynamics with quantum nuclei at $T = 90$~K are also shown (TRPMD). Dotted lines correspond to exponential fits to the correlation functions at long times.
}
\label{fig:hbcorrs}
\end{figure}

To understand how NQEs increase the rotational disorder of ammonia molecules,
we first quantify their impact on hydrogen bond dynamics within the \aacrystal~co-crystal, Fig.~\ref{fig:hbcorrs}.
In hydrogen bonded systems, NQEs can promote or obstruct the diffusional motion of molecules by strengthening or weakening the hydrogen bond network~\cite{ceriotti2016nuclear, habershon2009competing}. 
For instance, quantum delocalization of the nuclei along the hydrogen bonding direction results in stronger hydrogen bonds and slower dynamics while delocalization along the perpendicular directions results in weaker hydrogen bonds and faster dynamics~\cite{ceriotti2016nuclear, habershon2009competing}.   
%

%
We probe NQEs on both types of hydrogen bonds in the \aacrystal~co-crystal.
Classical hydrogen bonding time correlation functions at 30 and 90~K exhibit similar behavior for \chn~hydrogen bonds.
The \chn~hydrogen bond correlations undergo a slight decay at short times followed by constant correlations.   
The quantum correlation functions show a similar decay, although the magnitude of the short-time decay is slightly larger due to increased positional disorder of the quantum nuclei.   
The overall lack of decay in the classical and quantum hydrogen bonding correlations suggest that \chn~hydrogen bonds do not undergo significant breakage and reformation. 
This lack of hydrogen bond dynamics is consistent with the lack of acetylene rotations.
For the \nhpi~hydrogen bonds, both classical and quantum correlations decay on a finite timescale and resemble the decay of the N-H bond orientational correlations, suggesting that \nhpi~hydrogen bonds break and reform as ammonia molecules rotate. 
With quantum nuclei, the correlations of the \nhpi~hydrogen bonds decay faster than the classical correlations.
Fitting to an exponential decay yields a classical H-bond correlation time of 0.75~ps and a quantum correlation time of 0.28~ps. 
The faster hydrogen bond dynamics in the quantum system can be tied to the faster reorientational motion of the ammonia molecules as a consequence of NQEs. 

\subsection{Nuclear Quantum Effects on Rotational Thermodynamics}

\begin{figure}[tb]
\begin{center}
\includegraphics[width=0.30\textwidth]{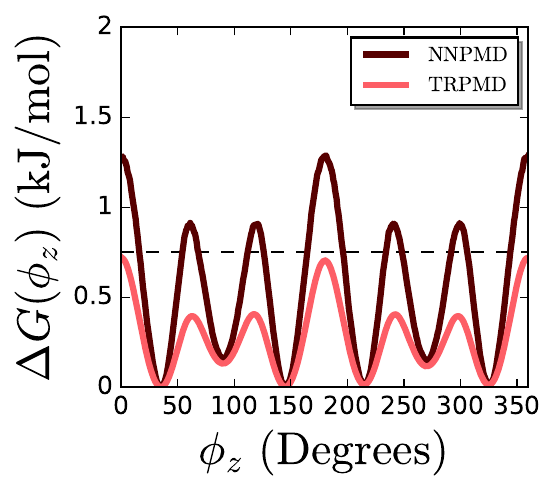}
\end{center}
\caption
{
Total free energy surface averaged over all ammonia molecules. Both classical (NNPMD) and quantum (TRPMD) free energy surfaces at $T = 90$~K are shown. The dotted line indicates the thermal energy, \kT.  
} 
\label{fig:fes}
\end{figure}

\begin{figure}[tb]
\begin{center}
\includegraphics[width=0.48\textwidth]{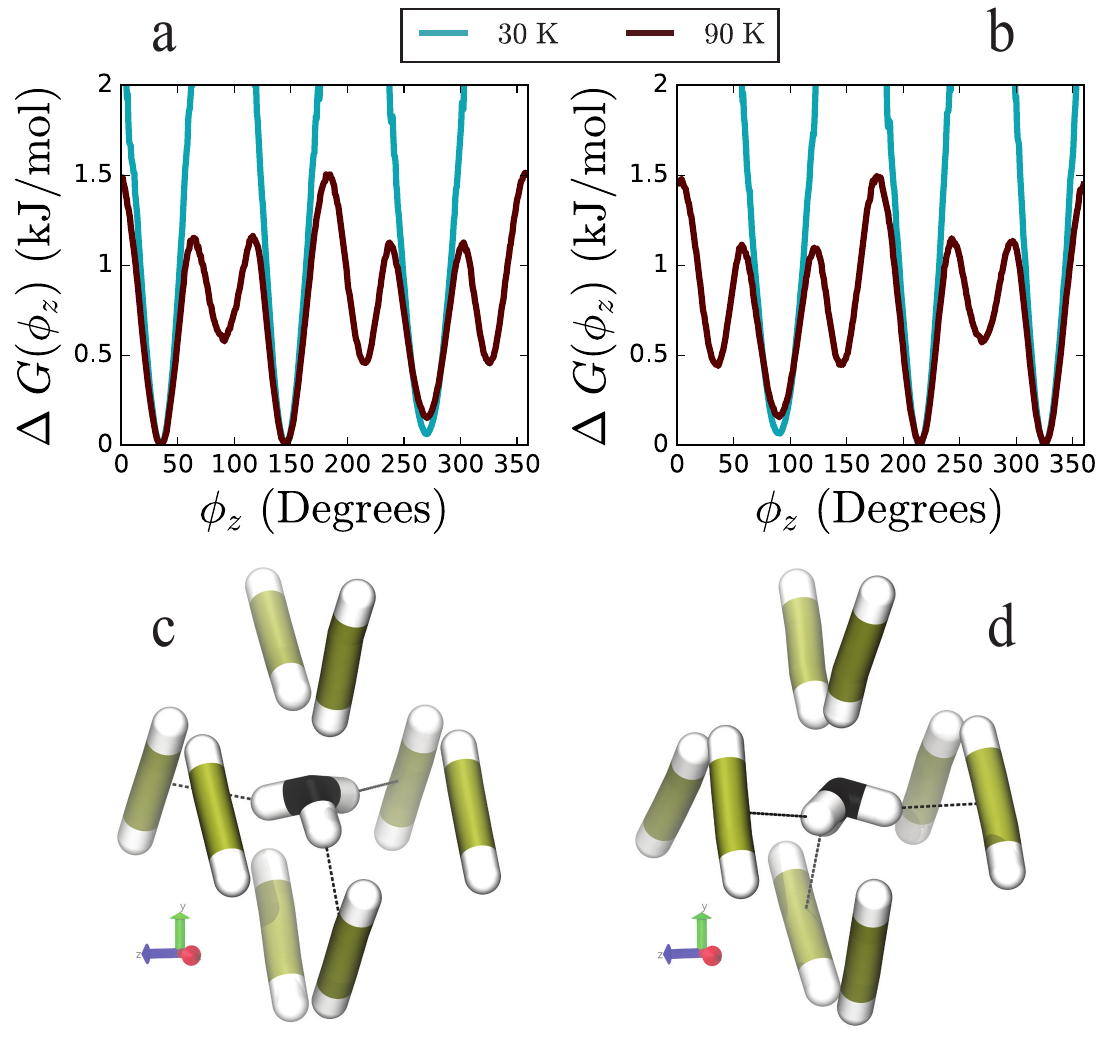}
\end{center}
\caption
{
Classical free energy surface as computed for the two types of orientations,  (a) Type~I and  (b) Type~II, of ammonia molecules within the \aacrystal~co-crystal at the two temperatures studied here. Snapshots of (c) Type~I and (d) Type~II orientations
are also shown, with \nhpi~hydrogen bonds shown as black dashed lines.
}
\label{fig:fesep}
\end{figure}

%
We can rationalize the faster reorientational dynamics of ammonia molecules in the quantum system by quantifying NQEs on thermodynamics.
NQEs can affect dynamics by decreasing relevant free energy barriers or nuclear tunneling through free energy barriers.    
We define a one dimensional free energy, $\Delta G(\phi_z)$, describing the rotations of ammonia as  
\begin{equation}
\Delta G(\phi_z) = -\kT \ln P(\phi_z),
\end{equation}
where $P(\phi_z)$ is the probability distribution of the azimuthal angle made by the N-H bond vector with the crystal $z$ axes,
shown in Fig.~\ref{fig:fes}
The free energy exhibits six minima and barriers when one might anticipate only three due to the symmetry of the ammonia molecule. 
The six minima in the free energy profile arise from the ability of
the ammonia molecules to adopt two hydrogen bonding orientations related to each other by a  $60\degree$ or $180 \degree$ rotation around the principle axis and a small rotation around a perpendicular axis, Fig.~\ref{fig:fesep}c,d.
We label these orientations as Type~I and Type~II, although apart from the rotation angle these two conformations are completely equivalent.   
At 30~K, when ammonia molecules do not rotate, the free energy profile of the two conformations
can be separated, Fig.~\ref{fig:fesep}.
The resulting free energy profiles for Type~I and Type~II ammonias are essentially the same
but offset by 90$\degree$.
At 90~K, however, the ammonia molecules can transition between the two orientations and a distinction between the two types cannot be made. 
As a result, $\Delta G(\phi_z)$ shows features for both types.
The six-fold symmetry in the total free energies in Fig.~\ref{fig:fes} is therefore a result of two free energy surfaces with three-fold symmetry, one for Type~I and one for Type~II.
%

Now that we understand the origins of the six-fold symmetry in $\Delta~G(\phi_z)$, we can compare the classical and quantum free energies, Fig.~\ref{fig:fes}.
In the classical systems, the free energy barriers to rotation are slightly larger than the thermal energy. 
The ammonia molecules transition between hydrogen bonding configurations through a jump diffusion process,
mainly with 120$\degree$ jumps, but also less probable 60$\degree$ jumps that transition the ammonia
molecules between Type~I and Type~II. 
Upon inclusion of NQEs, the barriers to rotation are significantly lowered, making the barriers easier to cross.
As a result, this lowering of free energy barriers by NQEs increases the dynamic rotational disorder in the quantum system.
We note that the transition between Type~I and Type~II orientations requires rotation around an axis perpendicular to $z$,
and the ammonia rotations in general are likely coupled to positional fluctuations of neighboring acetylene molecules~\cite{lynden1994translation}.
As a result, $\phi_z$ is not the only order parameter needed to fully describe the free energy landscape governing ammonia rotations, and the free energies presented here may not be sufficient to estimate kinetics. 
However, $\Delta G(\phi_z)$ adequately illustrates NQEs on free energy barriers to rotation; barriers are lowered by NQEs. 
NQEs will similarly lower free energy barriers on higher dimensional free energy landscapes involving additional order parameters. 
%

\section{Conclusion}

In this work, we used classical and quantum molecular dynamics simulations with neural network potentials to quantify NQEs on the structure and dynamics
of the \aacrystal~plastic co-crystal at Titan surface conditions. 
We find that NQEs increase structural and dynamic disorder within the co-crystal by weakening the hydrogen bond network, especially
those hydrogen bonds formed between ammonia molecules and the $\pi$-system of the acetylene molecules. 
This weakening of the hydrogen bond network is supported by shorter hydrogen bond lifetimes and lower ammonia rotational free energy barriers in the system with NQEs included. 
Because dynamic rotational disorder often softens the mechanical properties of plastic crystals, e.g. by reducing relevant elastic moduli,
we expect that NQEs will alter the mechanical response of plastic crystals on Titan's surface.
In addition, the weakening of hydrogen bonds holding the crystal together by NQEs may also promote
processes like dissolution at liquid-solid interfaces.
As a result, we expect that NQEs may need to be taken into account when predicting the properties of Titan cryominerals, especially those that may exist in a plastic crystal phase.
%

\begin{acknowledgements}
This work is supported by the National Aeronautics and Space Administration under grant number 80NSSC20K0609, issued through the NASA Exobiology Program.
We acknowledge the Office of Advanced Research Computing (OARC) at Rutgers,
The State University of New Jersey
for providing access to the Amarel and Caliburn clusters 
and associated research computing resources that have contributed to the results reported here.
This work used the Advanced Cyberinfrastructure Coordination Ecosystem: Services \& Support (ACCESS), formerly Extreme Science and Engineering Discovery Environment (XSEDE)~\cite{towns2014xsede}, which is supported by National Science Foundation grant number ACI-1548562. Specifically, this work used Stampede2 and Ranch at the Texas Advanced Computing Center through allocation TG-CHE210081.
We thank Harender Dhattarwal, Josiah Hall, Amanda Liyanaarachchi, and Olivia Martin for feedback on the manuscript.
\end{acknowledgements}

\section{Appendix A: Convergence of Training Neural Network Potentials and Comparison with Ab Initio Simulations}

\begin{figure}[tb]
\begin{center}
\includegraphics[width=0.48\textwidth]{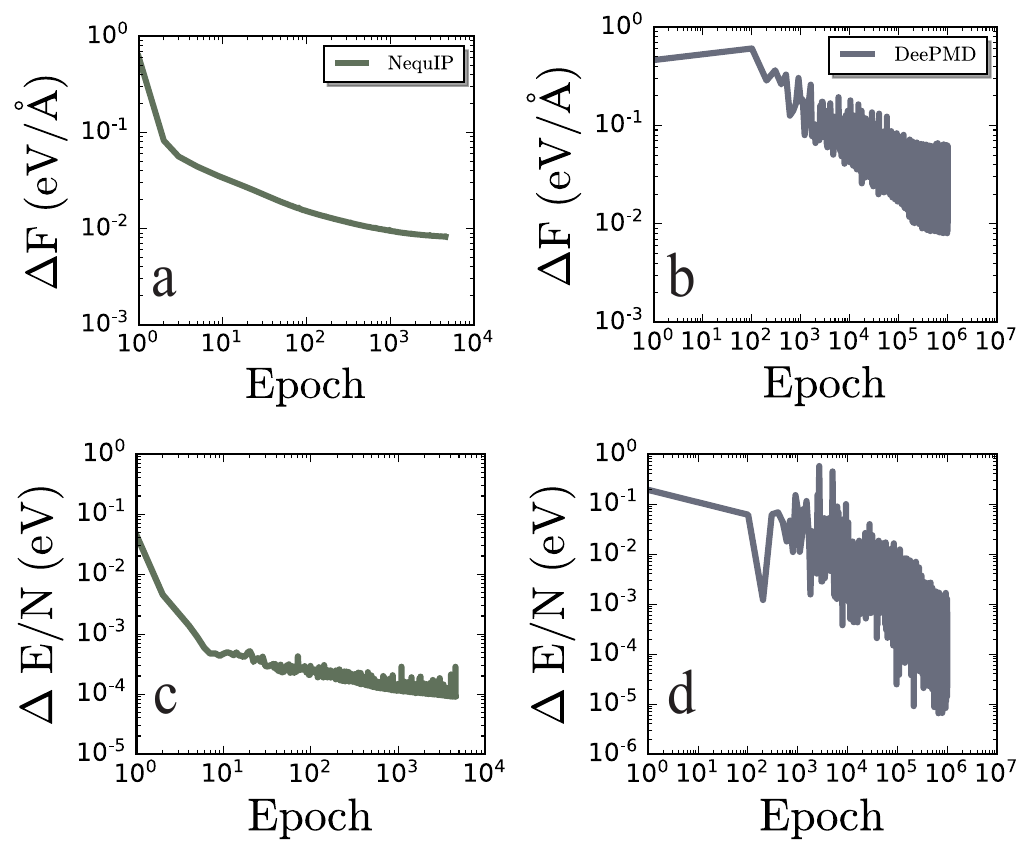}
\end{center}
\caption
{
Learning curves for the NequIP and DeePMD models showing the convergence of validation errors in (a) forces and (b) energies per atom on a log-log plot. NequIP errors are showing the mean absolute errors while the DeePMD curves are plotted using root mean square errors.   
}
\label{fig:energy_force}
\end{figure}

We assess the accuracy of the NequIP and DeePMD NN potentials by computing the learning curves and training errors,
as well as comparing the models against ab initio molecular dynamics (AIMD) simulations.
Note that the NequIP model was trained on 2000 configurations sampled from AIMD simulations at 90~K and 600~K, while the DeePMD model was trained on a larger data set consisting of 9000 configurations obtained from simulations at 30~K, 90~K, 200~K, and 600~K.
The errors in the energies and forces, quantified as the mean absolute errors for NequIP and root mean square errors for DeePMD, are shown in Fig.~\ref{fig:energy_force}.
The energy errors are converged to a value of $ 9.38 \times 10^{-5} ~{\rm eV/atom} $ for NequIP and $ 4.37 \times 10^{-5} ~{\rm eV/atom} $ for DeePMD, while the force errors respectively are $8.24 \times 10^{-3} ~{\rm eV/\AA} $  and $2.52 \times 10^{-2} ~{\rm eV/\AA}  $, which are expected to yield reasonably accurate results~\cite{batzner20223, pagotto2022predicting, yue2021short}.
 
\begin{figure}[tb]
\begin{center}
\includegraphics[width=0.48\textwidth]{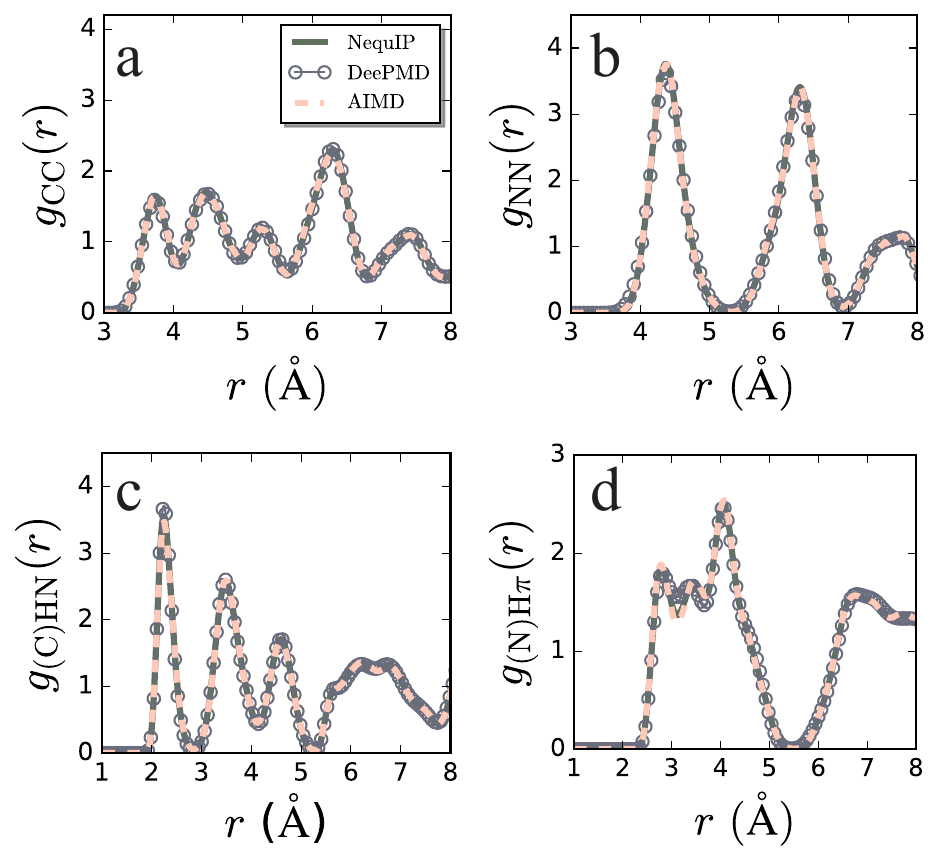}
\end{center}
\caption
{
Comparison of the site-site radial distribution functions (RDFs), $g(r)$, for sites within the \aacrystal~co-crystal, as obtained from ab initio simulations and from simulations using the NequIP and DeePMD NN models. 
}
\label{fig:nnaimdrdf}
\end{figure}

\begin{figure}[tb]
\begin{center}
\includegraphics[width=0.48\textwidth]{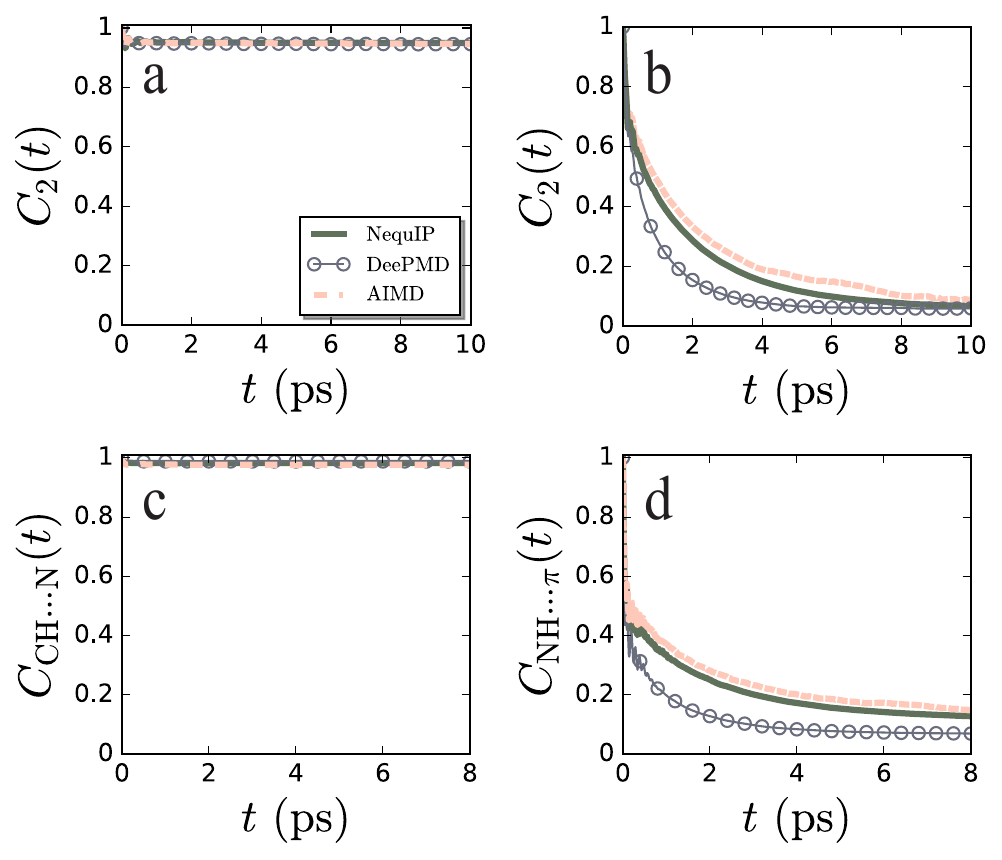}
\end{center}
\caption
{
Comparison of the $C_2(t)$ rotational correlations for (a) acetylene and (b) ammonia molecules as obtained from simulations using ab initio and NequIP and DeePMD NN potential at $T = 90$~K. 
Hydrogen bond time correlation functions for both (c) \chn~and (d) \nhpi~type hydrogen bonds are also shown. 
}
\label{fig:nnaimdcorrs}
\end{figure}

%
We also compare the structure produced by the NN potentials to that from the reference AIMD simulations. 
The partial structure for the carbon and nitrogen sites within the \aacrystal~computed using both NN potentials compares well with the predicted structure from AIMD simulations,
suggesting that both NN potentials adequately describe the potential energy surface of the \aacrystal~co-crystal, Fig.~\ref{fig:nnaimdrdf}a,b.
The RDFs relevant to \chn and \nhpi hydrogen bonding are also reproduced well by both NN potentials, Fig.~\ref{fig:nnaimdrdf}c,d.
However, DeePMD produces a slightly higher first minimum in the (N)H-$\pi$ RDF. 
The agreement between results from AIMD and NN potentials suggests that the NN models can describe the energetics of hydrogen bonding in the co-crystal.

As an even stricter test reflecting the accuracy of the NN potential, we compare the rotational and hydrogen bond dynamics of the ammonia molecules predicted by the NN potentials to AIMD results.
The rotational and hydrogen bond time correlation functions involving acetylene and \chn hydrogen bonds are reproduced well by both NN potentials, Fig.~\ref{fig:nnaimdcorrs}a,c.
The NequIP potential also accurately reproduces the rotational and hydrogen bond time correlation functions, Fig.~\ref{fig:nnaimdcorrs}b,d,
further suggesting that NequIP can accurately reproduce the energetics of hydrogen bonding within the co-crystal.
In contrast, the DeePMD potential produces rotational and hydrogen bond dynamics that are too fast.
This suggests that the DeePMD potential produces \nhpi hydrogen bonds that are too weak, resulting in classical dynamics that are similar to those produced by RPMD dynamics,
despite producing a more structured (N)H-$\pi$ than the quantum system.
The reason for the disagreement between the DeePMD and AIMD simulations is not immediately apparent,
but recent work on water suggests that deficiencies in describing many body effects within DeePMD can lead to such disagreements~\cite{zhai2023short}.

\bibliography{AAJCP}

\end{document}